\documentclass{article}
\usepackage{spconf,amsmath,graphicx}
\usepackage{epstopdf}
\usepackage{cite}
\usepackage{bm}
\usepackage{multirow}

\usepackage{booktabs}
\usepackage{amsmath}
\usepackage{color}
\usepackage{subcaption}
\usepackage{amssymb}
\usepackage[bottom]{footmisc}
\usepackage{soul}

\usepackage{url}

\usepackage{hyperref}
\usepackage{breakurl}
\usepackage{algorithmic}
\usepackage{algorithm}

\usepackage[flushleft]{threeparttable}

\newcommand{\inlineeqno}{\refstepcounter{equation}~~~~~~~~~~~~~~~~~~~\mbox{(\theequation)}}

\DeclareMathOperator*{\argmin}{arg\,min}
\newcommand\bx{\ensuremath{{\bm x}}}

\newcommand\bG{\ensuremath{{\bm G}}}
\newcommand\bB{\ensuremath{{\bm B}}}
\newcommand\bW{\ensuremath{{\bm W}}}
\newcommand\bL{\ensuremath{{\bm L}}}
\newcommand\bmu{\ensuremath{{\bm \mu}}}
\newcommand\bI{\ensuremath{{\bm I}}}
\newcommand\by{\ensuremath{{\bm y}}}
\newcommand\bA{\ensuremath{{\bm A}}}
\newcommand\bU{\ensuremath{{\bm U}}}
\newcommand\bPhi{\ensuremath{{\bm \Phi}}}
\newcommand\bDelta{\ensuremath{{\bm \Delta}}}

\title{Covariance Regularization for Probabilistic Linear Discriminant Analysis}
\name{Zhiyuan Peng$^{1,2,\dagger}$, Mingjie Shao$^{1,\dagger}$, Xuanji He$^2$, Xu Li$^3$, Tan Lee$^1$, Ke Ding$^2$, Guanglu Wan$^2$
\thanks{$^\dagger$Equal contribution. }
}

\address{
  $^1$Department of Electronic Engineering, The Chinese University of Hong Kong\\
  $^2$Meituan\\
  $^3$ARC Lab, Tencent PCG \\
 \begin{footnotesize}
 \texttt{jerrypeng1937@gmail.com,mingjieshao@cuhk.edu.hk}
 \end{footnotesize}\\
  \begin{footnotesize}
    \texttt{\{hexuanji,dingke02,wanguanglu\}@meituan.com,nelsonxli@tencent.com,tanlee@ee.cuhk.edu.hk}
 \end{footnotesize}
 }

\begin{document}
\ninept
\maketitle
\begin{abstract}
Probabilistic linear discriminant analysis (PLDA) is commonly used in speaker verification systems to score the similarity of speaker embeddings. Recent studies improved the performance of PLDA in domain-matched conditions by diagonalizing its covariance. We suspect such brutal pruning approach could eliminate its capacity in modeling dimension correlation of speaker embeddings, leading to inadequate performance with domain adaptation. This paper explores two alternative covariance regularization approaches, namely, interpolated PLDA and sparse PLDA, to tackle the problem. The interpolated PLDA incorporates the prior knowledge from cosine scoring to interpolate the covariance of PLDA. The sparse PLDA introduces a sparsity penalty to update the covariance. Experimental results demonstrate that both approaches outperform diagonal regularization noticeably with domain adaptation. In addition, in-domain data can be significantly reduced when training sparse PLDA for domain adaptation.

\end{abstract}

\section{Introduction}\label{sec:intro}

Automatic speaker verification (ASV) is the process of verifying whether a given speech utterance comes from a claimed speaker or not\cite{hansen2015humans}. A typical ASV system consists of a front-end embedding extractor and a back-end scoring model. The front-end extractor compresses the utterance into a fixed-dimensional vector representation, termed as speaker embedding, that captures speaker-related acoustic attributes. The back-end model measures the similarity between two sets of speaker embeddings and determines whether they represent the same person.

The use of neural networks as the front-end has shown great success in ASV\cite{snyder2017deep,desplanques2020ecapa}. With deep layers, complex neural blocks, and angular-based large margin penalties\cite{wang2018additive,wang2018cosface,deng2019arcface}, the front-end neural network can achieve the within-speaker compactness and increase the between-speaker discrepancy of speaker embeddings, thus leading to substantial performance improvements over the classical i-vector method\cite{dehak2010front}. Following the terminology in \cite{wang2022scoring}, the embeddings extracted from networks trained with margin penalties are termed large-margin embeddings in this paper.

As for back-end modeling, two methods are commonly used, namely, cosine scoring and probabilistic linear discriminant analysis (PLDA) scoring\cite{ioffe2006probabilistic} . Given two speaker embeddings after length normalization\cite{garcia2011analysis}, the cosine scoring computes their inner product as the similarity score. The method is intuitively straight-forward, parameter-free, and easy to implement and deploy. The PLDA back-end assumes the speaker embeddings extracted from the front-end are Gaussian distributed and controlled by a set of latent speaker factors. Both the training and the scoring processes follow rigorous theoretical analysis. The training of PLDA back-end demands the use of speaker labels. Compared to cosine scoring, PLDA scoring is expected to have better  discrimination power on speaker-related information in the embeddings.

However, empirical studies\cite{liu2019large,zhou2020dynamic} showed that the emergency of large-margin embeddings shifts the choice of back-end from theoretically appealing PLDA to simple cosine similarity measure, especially when the test data are collected under the domain-matched condition, i.e.,  a domain covered in the front-end training. When an SV system is deployed in the real world, it has to deal with diverse input data from a myriad of domains. Unexpected cross-domain problems such as channel mismatch, duration shift and time drift, seriously degrade the system performance\cite{li2022coral++}. In this regard, a common 
approach is to incorporate a small set of data from the target domain for fast back-end adaptation. PLDA, as a trainable model, is naturally suitable for back-end adaptation\cite{borgstrom2021unsupervised,borgstrom2020bayesian,lee2019coral+,bousquet2019robustness,villalba2014unsupervised}.

Although PLDA performs well with back-end adaptation, its inferiority against the simple cosine measure in the domain-matched condition needs to be addressed. Two very recent works\cite{peng2022unifying,wang2022scoring} provide both theoretical and empirical supports that regularization is needed for the covariance of PLDA . In \cite{wang2022scoring}, it was hypothesized that the within-speaker compactness of large-margin embeddings
makes the modeling of within-speaker variability in PLDA not essential. The other study\cite{peng2022unifying} suggested that large-margin embeddings are likely to be dimension independent, which should be taken into PLDA training. Both of these works considered to diagonalize covariance of PLDA in training, leading to diagonal PLDA.

Here we suspect that it could be too brutal to prune all the off-diagonal elements of the covariance. The diagonal PLDA fails to capture dimension correlation of speaker embeddings under domain-mismatched conditions and performs inadequately with back-end adaptation\cite{peng2022unifying}. Alternative regularization approaches should be explored to limit instead of eliminating the capacity of PLDA in modeling dimension correlation. 

This paper proposes interpolated PLDA (I-PLDA) and sparse PLDA (S-PLDA) to address the issue aforementioned. The I-PLDA, inspired by the unsupervised bayes PLDA\cite{borgstrom2020bayesian}, incorporates prior knowledge from cosine scoring to interpolate the covariance in PLDA training. The S-PLDA adjusts the objective of the M-step in PLDA training by introducing a sparsity penalty, which encourages sparse inverse covariance estimation. 
The efficacy of the proposed approaches is experimentally validated on VoxCeleb\cite{nagrani2017voxceleb} and CNCeleb\cite{li2022cn}.

\section{Why covariance regularization ?\label{sec:why}}
PLDA, despite being theoretically appealing, was shown inferior to straight-forward cosine scoring in many empirical studies\cite{liu2019large,zhou2020dynamic}. Our recent work\cite{peng2022unifying} gave a theoretical account showing that cosine scoring is inherently a variant of PLDA. This suggested that the non-Gaussian nature of speaker embeddings does not necessarily lead to the inferiority of PLDA. We advocated that cosine scoring demands more stringent constraints on the embeddings such that:
\begin{enumerate}
    \item (\textbf{dim-indep}) Individual dimensions of speaker embeddings are uncorrelated or independent;
    \item In addition to 1, all dimensions share the same variance value.
\end{enumerate}
It was experimentally verified that the dim-indep assumption contributes significantly to the performance gap between the two scoring models. The assumption suggests diagonally regularizing the covariance of PLDA in its EM training. Empirical results showed that substantial improvement could be achieved on SV with PLDA back-end when covariance regularization was applied\cite{wang2022scoring}.

Insufficient speaker instances may be another potential issue for PLDA training. Recall the EM training algorithm, as described in Algorithm \ref{alg:plda_em}. According to Eq.~\eqref{eq:update_plda_B} in the M-step, estimation of the between-speaker covariance $\bB ^{-1}$ takes $M$ per-speaker embeddings $y_m$ into accumulation. $\bB^{-1}$ is of dimension $D\times D$, where $D$ denotes the embedding dimension. $\bB^{-1}$ contains $D(D+1)/2$ free parameters (halved due to its symmetry property)\cite{sizov2014unifying} to be estimated. Therefore, for high-dimension embedding, the number of free parameters could be undesirably large. On the other hand, speakers involved in training could be limited. In this context, the estimation of $\bB^{-1}$ in Eq.~\eqref{eq:update_plda_B} could be unreliable to a certain extent. Consequently, regularization on  $B^{-1}$ is necessary, especially when adapting PLDA on limited in-domain data.

\begin{algorithm}[H]
\caption{E-M training of two-covariance PLDA}\label{alg:plda_em}
\begin{algorithmic}
 \STATE \textbf{Notation}:
 \STATE \hspace{0.3cm} number of speakers $M$
 \STATE \hspace{0.3cm} number of utterances $N$
 \STATE \hspace{0.3cm} number of utterances from the $m$-th speaker $n_m$
 \STATE \hspace{0.3cm} global mean $\bmu$
 \STATE \hspace{0.3cm} between-speaker covariance $B^{-1}$
 \STATE \hspace{0.3cm} within-speaker covariance $W^{-1}$
 \STATE \textbf{Input}: per-utterance embeddings $\mathcal{X} =\{\bx_{m,n}\}_{1,1}^{M, n_m}$
 \STATE \textbf{Initialization}: $\bB=\bW=\bI, \bmu = \mathbf{0}$
 \REPEAT
 \STATE \textbf{(E-step):} Infer the latent variable $\by_m | \mathcal{X}$
 \STATE \hspace{0.3cm} $\bL_m = \bB + n_m \bW$
 \STATE \hspace{0.3cm} $\by_m |\mathcal{X} \sim \mathcal{N} (\bL_m^{-1} (\bB\bmu + \bW \sum_{n=1}^{n_m}x_{m,n}), \bL_m^{-1})$
 
 \STATE \textbf{(M-step):} Update model parameters
 \STATE \hspace{0.3cm} $\bmu = \frac{1}{M} \sum_m \mathbb{E}[\by_m|\mathcal{X}]$
 \STATE \hspace{0.3cm} {\color{blue}$
     \bB^{-1} = \frac{1}{M} \sum_m \mathbb{E} [ \by_m \by_m^T| \mathcal{X} ] - \bmu \bmu^T \inlineeqno \label{eq:update_plda_B}$}
 \STATE \hspace{0.2cm} $\bW^{-1} = \frac{1}{N} \sum_m \sum_n \mathbb{E} [ (\by_m-\bx_{m,n})(\by_m-\bx_{m,n})^T | \mathcal{X} ]$
 \UNTIL{Convergence}\;
 \STATE \textbf{Return} $\bB, \bW, \bmu$

\end{algorithmic}
\label{alg1}
\end{algorithm}

\section{PLDA with covariance regularization}
We explore and compare three regularization methodologies that are expected to benefit PLDA. In accordance with the dim-indep assumption, the regularization should not only prune the free parameters but also constrain the covariance such that non-zero values concentrate along the diagonal. Let $\bG = \frac{1}{M}\sum_m \mathbb{E} [ \by_m \by_m^T| \mathcal{X} ] - \bmu \bmu^T$. $\bG$(the right-hand side of Eq.~\eqref{eq:update_plda_B}). $\bG$ is regarded as the statistical estimation of $\bB^{-1}$ without regularization.

\subsection{Diagonal PLDA}
Following the dim-indep assumption, a straight-forward regularization approach is to prune all off-diagonal elements in $\bG$ when updating $\bB^{-1}$.
It was noted empirically in \cite{wang2022scoring} that PLDA benefits from covariance diagonalization, especially under domain-matched test conditions. However, such assumption is considered too strict and strong in practice. In a real scenario, the front-end embedding extractor may suffer from serious domain mismatch, which leads to the violation of this assumption. This results in the inferiority of covariance diagonalization in domain adaptation.

\subsection{Interpolated PLDA}
From the bayesian perspective, a common practice for model regularization is by imposing a prior distribution on the parameters to be estimated. Similar to  VB-MAP in \cite{borgstrom2021unsupervised}, a Wishart prior is assumed for $\bB$, e.g., $p(\bB) = \mathcal{W} (\frac{1}{\gamma M}\bG_0, \gamma M)$ . Accordingly, $\bB^{-1}$ can be updated by maximum \textit{a posterior} (MAP), i.e.,
\begin{equation}
    \bB^{-1} = \frac{1}{1+\gamma} \bG + \frac{\gamma}{1+\gamma} \bG_0, \label{eq:interpolation_map}
\end{equation}
where $\gamma$ and $\bG_0$ denote the hyper-parameters of the Wishart prior. Simply speaking, the update is given by a linear interpolation of estimation from observed data and prior knowledge. As shown in \cite{peng2022unifying}, cosine scoring is equivalent to PLDA scoring if and only if $\bB=\bW=\bI$. Therefore, let us set $\bG_0 = \bI$ to impose this prior into PLDA training. As empirically suggested in \cite{borgstrom2021unsupervised}, we let $\gamma = 2$.

\subsection{Sparse PLDA}
The third approach is by imposing sparsity on the inverse covariance matrix, e.g., adding $\ell_1$ penalty in the update of $\bB$. In this way, elements of $\bB$ are propelled towards zero and high sparsity is encouraged. The formulation is given as follows,
\begin{equation}
    \argmin_{\bB \succeq \mathbf{0}} ||\bB - \bG^{-1}||_2^2 + \lambda ||\bB||_1 \label{eq:admm_origin_obj},
\end{equation}
where $\lambda$ controls the degree of penalty. This problem can be solved by alternating direction method of multipliers (ADMM). Specifically, Eq.~\eqref{eq:admm_origin_obj} can be written as,
\begin{align}
    \argmin_{\bB \succeq \mathbf{0}} & \frac{1}{2}||\bB - \bG^{-1}||_2^2 + \lambda ||\bA||_1 \nonumber \\
    {s.t.} & \bA = \bB \label{admm_convert_obj}
\end{align}
The augmented Lagrangian form can be derived as
\begin{align}
    L (\bB, \bA, \bPhi) = & \frac{1}{2}||\bB-\bG^{-1}||_2^2  + <\bPhi, \bA - \bB> \nonumber \\
    & + \frac{\beta}{2} ||\bA-\bB||_F^2 + \lambda ||\bA||_1,
\end{align}
where $\bPhi$ is an auxiliary variable introduced for dual update. $\beta$ guarantees the equality between $\bA$ and $\bB$. The variables $\bB, \bA, \bPhi$ are consecutively updated as follows,
\begin{align}
    \bB^{k+1} &= \argmin_{\bB \succeq 0} L (\bB, \bA^k, \bPhi^k) \label{eq:update_B}\\
    \bA^{k+1} &= \argmin_{\bA} L (\bB^{k+1}, \bA, \bPhi^k) \label{eq:update_A} \\
    \bPhi^{k+1} &= \bPhi^k + \beta (\bA^{k+1} - \bB^{k+1})
\end{align}
Eq.~\eqref{eq:update_B} can be solved by the iterative projected gradient descent method \cite{beck2017first}. Eq.~\eqref{eq:update_A} has a closed-form solution that is achieved by soft thresholding \cite{beck2017first}.
Details are given in Algorithm \ref{alg:admm}. 
\begin{algorithm}[H]
\caption{Sparse regularization of $\bB$}\label{alg:admm}
\begin{algorithmic}
 \STATE \textbf{Input}: $\bG$, $\lambda$,   $\beta$, $\epsilon$
 \STATE \textbf{Initialization}: $\bB^0$ estimated from the last M-step, $\bA^0=\bB^0, \bPhi^0=\mathbf{0}$
 \REPEAT
 \STATE \hspace{0.01cm} Solve Eq.~(6) by projected gradient descent
 \STATE \hspace{0.01cm} \textbf{repeat}
 \STATE \hspace{0.3cm} $\nabla L_\bB(\bB, \bA, \bPhi) = \bB - \bG^{-1}-\bPhi + \beta (\bB-\bA)$
 \STATE \hspace{0.3cm} $\bB^* = \bB^k - 1/(1+\beta) \cdot\nabla L_\bB(\bB^k, \bA^k, \bPhi^k)$
 \STATE \hspace{0.3cm} Compute EVD: $\bU\bDelta\bU^T = \bB^*$
 \STATE \hspace{0.3cm} Compute $\hat{\bDelta}$: $\hat{\bDelta}_{ii} = max(\bDelta_{ii}, 0)$
 \STATE \hspace{0.3cm} $\bB^{k+1} = \bU \hat{\bDelta} \bU^T$
 \STATE \hspace{0.01cm} \textbf{until} $||\bB^{k+1}-\bB^k||_F < \epsilon $
 \STATE \hspace{0.01cm} Solve Eq.~(7) by soft thresholding
 
 \STATE \hspace{0.3cm} $\bA^* =  \bB^{k+1} -  \bPhi^k/\beta$
 \STATE \hspace{0.3cm} $\bA^{k+1}_{ij} = \max(0, |\bA_{ij}^*| - \lambda/\beta)\cdot  \text{sgn}(\bA^*_{ij})$
 \STATE \hspace{0.01cm} Solve Eq.~(8): $\bPhi^{k+1} = \bPhi^k + \beta (\bA^{k+1}-\bB^{k+1})$
 \UNTIL{ $||\bA^{k+1}-\bB^{k+1}||_F < \epsilon$ }\;
 \STATE \textbf{Return} $\bB$
\end{algorithmic}
\label{alg:admm}
\end{algorithm}
Our experimental results suggest the hyper-parameters set as follows: $\lambda =1e^{-3}$, $\beta=0.1$, $\epsilon=1e^{-6}$.

\section{Experimental Configuration}

\subsection{Front-end\label{sec:front-end}}
ECAPA-TDNN\cite{desplanques2020ecapa} is adopted as the front-end embedding extractor. It consists of three squeeze-excitation (SE) blocks, multi-scale Res2Net feature maps, multi-layer feature aggregation, and channel-dependent attentive statistics pooling. The
channel size and the bottleneck dimension in the SE blocks are set to 1024 and 256, respectively. The input features are 80-dim filter-bank features processed by voice activity detection and mean normalization over a sliding window of up to 3 seconds. AAM-softmax loss\cite{liu2019large} is applied in training with the margin and the scale of 0.2 and 30 respectively. After training, 192-dimensional speaker embeddings can be extracted from the last linear layer.

\subsection{Back-end\label{sec:back-end}}
All back-end models are implemented by PyKaldi\cite{can2018pykaldi}, including the conventional two-covariance PLDA, the diagonal PLDA (D-PLDA), the interpolated PLDA (I-PLDA) and the sparse PLDA (S-PLDA). For comparison, we implement the unsupervised bayes PLDA (UB-PLDA) that trains a PLDA model with variational bayes maximum \textit{a posterior} (VB-MAP)\cite{borgstrom2021unsupervised}. In UB-PLDA, pseudo speaker assignments are randomly initialized from Dirichlet distribution. They can also be initialized from ground-truth speaker labels. We denote the bayes PLDA initialized in this way as supervised bayes PLDA (SB-PLDA).
The codes related to bayes PLDA are available on GitHub\footnote{https://github.com/JerryPeng21cuhk/bayes-plda}. 

\subsection{Data\label{sec:data}}
The front-end embedding extractor is trained on the training data of VoxCeleb1\cite{nagrani2017voxceleb} and VoxCeleb2\cite{chung2018voxceleb2}, which contain 1,240,651 speech clips from 7205 celebrities. The clips are mostly interview speech utterances collected from YouTube. Data augmentation is applied with speed perturbation, additive noise, and reverberation, following the recipe\footnote{https://github.com/kaldi-asr/kaldi/blob/master/egs/voxceleb/v2/run.sh} in Kaldi. The noise datasets, including MUSAN\cite{snyder2015musan} and RIRS NOISES\cite{ko2017study}, are adopted in the data augmentation.

The original test set of VoxCeleb1 (Vox1-o) is exploited to evaluate the back-end models under the domain-matched condition. Meanwhile, the CNCeleb1 (CN1)\cite{li2022cn} dataset is leveraged for back-end adaptation under domain-mismatched conditions. It includes 800 speakers with 111,259 speech clips for training and 200 speakers with 18,224 speech clips for evaluation. The data are collected from 11 diverse genres that differ distinctively from the VoxCeleb datasets.
Results are reported in terms of equal error rate (EER) and the minimum normalized detection cost function (minDCF) at $P_\text{target}=10^{-2}$ and $C_\text{FA}=C_\text{Miss}= 1$.

\section{Result and Analysis}
\subsection{Which covariance to regularize ?}
We first investigate which covariance of PLDA needs regularization. Results are given as Table \ref{tab:exp1}. The evaluation condition with Vox1-o is ideal where the data for training the front-end is similar to the test data, i.e., the domains are matched. Under such conditions, regularization on both $W^{-1}$ and $B^{-1}$ shows the best performance for D-PLDA and I-PLDA. For S-PLDA, the choice of covariance for regularization has a minor effect on its performance. Nevertheless, all of the three PLDA variants outperform the conventional PLDA model by a noticeable margin.

Experiments on CN1 reveal more realistic evaluation results where diverse test conditions cannot be covered by
the finite amount of data for front-end training. In such scenario, the back-ends are re-trained by fast adaptation with CN1 training data. As shown in Table \ref{tab:exp1}, the experimental results on CN1 suggest that regularization should be carried out with $B^{-1}$ only. 
The primary practical concern of developing an ASV system is its robustness toward domain-mismatched conditions. From this perspective, regularizing only $B^{-1}$ may be preferable. The cost is a minor performance drop in the domain-matched case. In the following experiments, regularization is applied to $B^{-1}$ only.

\begin{table}[htbp]
\caption{Evaluation in terms of EER/minDCF for back-ends. Regularization can be applied on $W^{-1}$, $B^{-1}$ or both. Back-ends except non-parametric cosine scoring (Cos) are trained and evaluated on Vox1 or CN1.\label{tab:exp1}}
\centering
\begin{tabular}{|l|c|c|c|c|}
\hline
\multicolumn{1}{|l|}{}                        & $W^{-1}$ & $B^{-1}$ & Vox1-o & CN1\\ \hline
\hline
\multicolumn{1}{|l|}{\multirow{3}{*}{D-PLDA}} & \checkmark          &           & 1.39/0.142 & 14.80/0.572  \\
\multicolumn{1}{|l|}{}                        &              & \checkmark           &  2.22/0.189 & \textbf{11.77}/\textbf{0.543} \\
\multicolumn{1}{|l|}{}                        & \checkmark          & \checkmark           & \textbf{0.97}/\textbf{0.107} & 13.15/0.607  \\ \hline
\multicolumn{1}{|l|}{\multirow{3}{*}{I-PLDA}} & \checkmark          &               & 1.09/0.147 & 13.64/0.589 \\
\multicolumn{1}{|l|}{}                        &              & \checkmark            & 1.24/0.173 & \textbf{9.44}/\textbf{0.511}\\
\multicolumn{1}{|l|}{}                        & \checkmark          & \checkmark           & \textbf{1.00}/\textbf{0.139} &  11.93/0.570  \\ \hline
\multicolumn{1}{|l|}{\multirow{3}{*}{S-PLDA}} & \checkmark          &               & \textbf{0.89}/0.108 & 13.01/0.538 \\
\multicolumn{1}{|l|}{}                        &              & \checkmark             & 0.98/0.119 & \textbf{10.51}/\textbf{0.528}\\
\multicolumn{1}{|l|}{}                        & \checkmark          & \checkmark           & 0.95/\textbf{0.106} &  12.53/0.577   \\ \hline
PLDA                                          &      -        &        -        & 1.83/0.205 & 10.25/0.536  \\ \hline
Cos                                        &      -        &     -          & 1.04/0.113 & 12.02/0.577  \\ \hline
\end{tabular}
\end{table}

\subsection{Adaptation under the multi-genre condition}
\begin{figure*}[htbp]
    \centering
    \includegraphics[width=\linewidth]{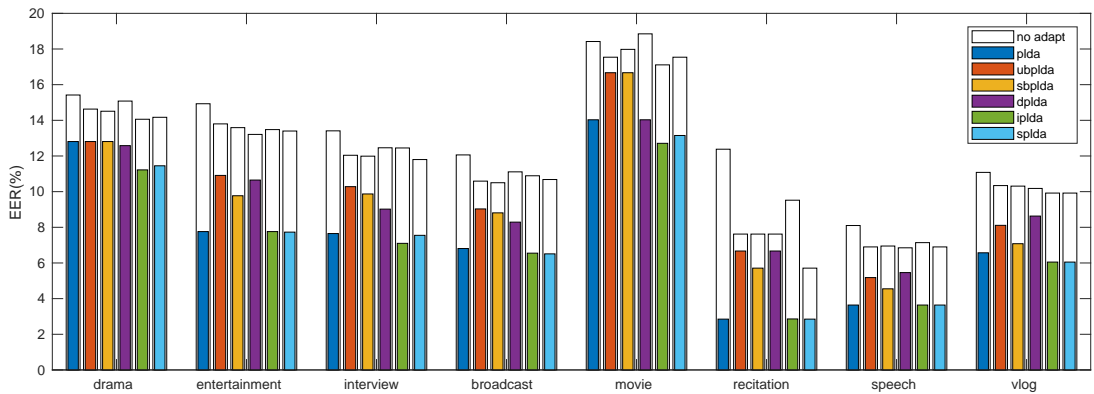}
    \caption{Analysis of back-end performance under the multi-genre condition in CN1}
    \label{fig:exp2}
\end{figure*}

Test speech in ASV could come from diverse domains. Performance evaluation should be carried out with multi-genre speech data. In this section, the back-ends are evaluated with 8 selected genres from CN1 as shown in Fig.\ref{fig:exp2}. The back-ends are trained on training data from either out-of-domain Vox1 (\textit{no adapt}) or the in-domain CN1. All back-ends show consistent improvements with adaptation. Admittedly, PLDA with abundant in-domain data could achieve comparable performance to I-PLDA and S-PLDA. But its robustness on \textit{no adapt} is problematic. On the contrary, D-PLDA slightly outperforms PLDA under \textit{no-adapt} for most genres. Its adaptation performance is far from satisfactory, probably because the dimensional correlation of speaker embeddings is not modeled.

UB-PLDA and SB-PLDA are evaluated, as illustrated in Section \ref{sec:back-end}. UB-PLDA in general shows a slightly worse performance than PLDA for adaptation, as it is unsupervisedly trained. SB-PLDA with pseudo speaker assignments initialized from ground-truth labels could gain slight improvement over UB-PLDA for adaptation. If the pseudo speaker assignments are fixed as the ground-truth labels, SB-PLDA would degrade to I-PLDA with both between-speaker and within-speaker covariances regularized. In this context, the performance gap between UB-PLDA and I-PLDA could be largely attributed to the use of speaker labels.

\subsection{Adaptation with limited data}
When abundant in-domain data are available for adaptation, fine-tuning the heavy front-end may be more beneficial than adapting the back-end. Fast back-end adaptation with limited in-domain data would be meaningful and desired in practical applications. Table \ref{tab:exp3} gives the performance of the back-ends when in-domain data are limited. The benefit of covariance regularization is significant. With a subset of in-domain data from 10 speakers, S-PLDA is able to reach a comparable performance to PLDA trained with data from 100 speakers. UB-PLDA and SB-PLDA can outperform PLDA when very limited speakers are available, but the performance remains almost unchanged as the amount of training data increases. For the scenario where a small amount of in-domain labeled data are available, we would recommend S-PLDA for domain adaptation.

\begin{table}[htbp]
\caption{EER of back-ends evaluated on the CN1 interview genre. Random subsets of speakers from CN1 training data are selected for back-end adaptation. The experiment is repeated for five times and the EERs reported are averaged.\label{tab:exp3}}
\resizebox{0.45\textwidth}{!}{
\begin{tabular}{|l|l|l|l|l|l|l|}
\hline
 \#spk  & 10            & 20            & 40            & 60            & 100           & 200           \\ \hline
 \hline
PLDA    & 20.77         & 15.12         & 11.46         & 10.34         & 9.27          & 8.28          \\
UB-PLDA & 10.21         & 10.15         & 10.14         & 10.13         & 10.10         & 10.14         \\
SB-PLDA & 10.01         & 9.83          & 9.76          & 9.78          & 9.82          & 9.80          \\
D-PLDA  & 10.95         & 9.91          & 9.73          & 9.47          & 9.29          & 9.09          \\
I-PLDA  & 14.20         & 12.39         & 10.25         & 9.45          & 8.49          & \textbf{7.56} \\
S-PLDA  & \textbf{9.10} & \textbf{8.55} & \textbf{8.35} & \textbf{8.27} & \textbf{8.14} & 8.08          \\ \hline
\end{tabular}}
\end{table}

\section{Conclusion}
Diagonal regularization of PLDA covariance (D-PLDA), despite its success under domain-matched conditions, performs inadequately on back-end adaptation in ASV systems.
To address this issue, the present study investigates more sophisticated approaches to covariance regularization, namely, interpolated PLDA (I-PLDA) and sparse PLDA (S-PLDA). There are three key findings. First, with domain adaptation, regularizing only the between-speaker covariance shows consistent benefit in system performance. Second, evaluation of adaptation on multi-genre data reveals that both I-PLDA and S-PLDA not only outperform D-PLDA but also attain comparable performance to the conventional PLDA. Third, S-PLDA significantly reduces the training data requirement for domain adaptation, making it a preferable technique of fast adaptation with limited in-domain data.

\bibliographystyle{IEEEtran}
\begin{footnotesize}
\bibliography{Template}
\end{footnotesize}

\end{document}